\documentclass[final,pra,aps,twocolumn,showpacs]{revtex4}
\usepackage{amsfonts}
\usepackage{amsmath}
\usepackage{amssymb}
\usepackage[english]{babel}
\usepackage[pdftex]{graphicx}
\usepackage{capt-of}
\usepackage{indentfirst}

\newcommand{\ud}{\mathrm{d}}

\begin{document}
\title[Multiphoton population dynamics in a three-level lambda system]{Multiphoton population dynamics in a three-level lambda system and the related light-scattering spectrum}
\author{M~Berent}
\email{mberent@amu.edu.pl}
\author{R~Parzy\'nski}
\affiliation{Faculty of Physics, Adam Mickiewicz University, Umultowska 85, 61-614 Pozna\'n, Poland}
\date{\today}

\begin{abstract}
We generalize our analytic approach (Berent~M and Parzy\'nski~R 2009  {\it Phys.~Rev.}~A~{\bf{80}} 033834) to include the effect of strong depopulation of the initial state when considering the multiphoton population transfer and scattering of low-frequency light by a 3-level system in the lambda-type configuration. We discuss the quality of the approximations made, i.e., the adiabatic and generalized rotating-wave approximations.
\end{abstract}

\pacs{42.50.Hz, 32.80.Rm, 33.80.Rv, 82.53.Kp}

\maketitle
\section{Introduction}
\begin{figure}[!ht]
\centering
\includegraphics[width=0.4\textwidth]{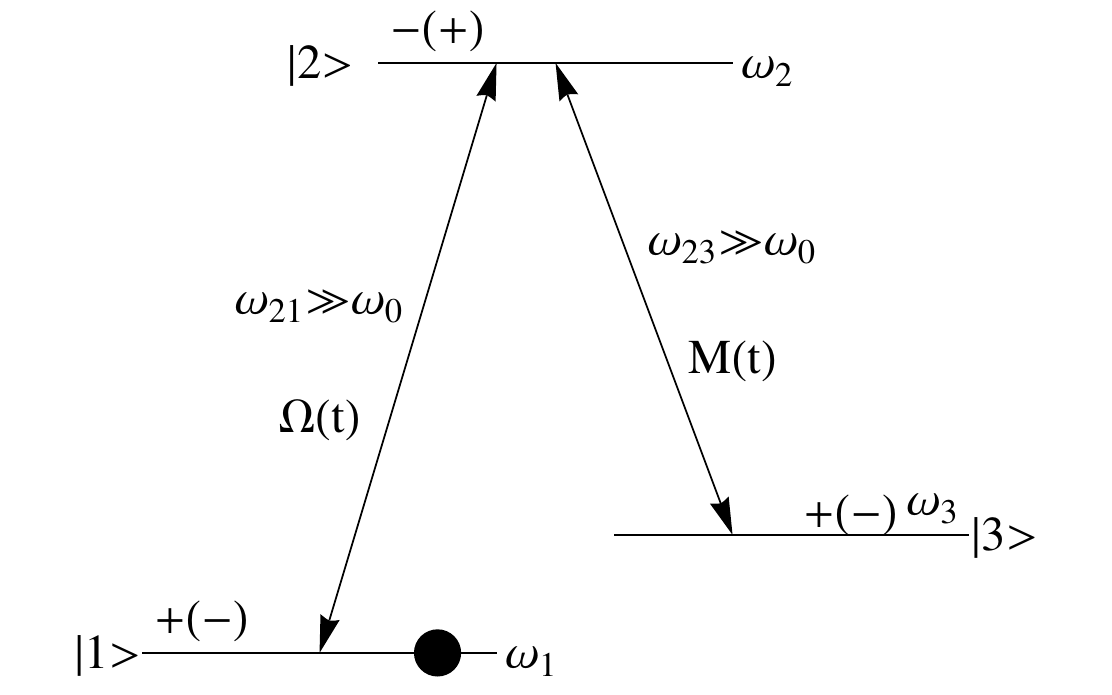}
\caption{The system under study with the separation frequencies $\omega_{21}$ and $\omega_{23}$ much higher than the laser frequency $\omega_0$. $\Omega(t)$ and $M(t)$ are the instantaneous Rabi frequencies for the electric-dipole 1-2 and 2-3 couplings, respectively.}\label{fig:lambda}
\end{figure}

\indent In an analytic approach \cite{parz}, we have recently described the spectrum of light scattered by a three-level lambda-type system under two conditions. One condition was that the incident-light frequency was much lower than the separation frequencies between the states situated in the ends of the arms of lambda configuration (Figure \ref{fig:lambda}). Thus, the transitions in the arms had multiphoton character. The other condition was that of weak depopulation of the initial state placed at the bottom of one arm (the black dot in Figure~\ref{fig:lambda}). As long as the latter condition was fulfilled, our analytic approach succeeded in explaining all details of the spectra found by numerical integration of the Schr\"odinger equation (number of peaks, their internal structures, positions, relative heights and sensitivity to the light-pulse shape and strength). Taking the wavefunction of the system as $\sum_{j=1}^3 b_j|j\rangle$, the basis of our approach in \cite{parz} was a set of two coupled nonlinear equations for the ratios $r = b_2/b_1$ and $\rho = b_3/b_1$ of the Schr\"odinger population amplitudes $b_j$. This set, found as
\begin{subequations}
\label{eq:ric}
\begin{eqnarray}
i\,\dot{r} &=& (r^2-1)\Omega(t) + \omega_{21}r-M(t)\rho\,,\label{eq:ric1}\\
i\,\dot{\rho} &=& (\omega_{31}+\Omega(t)r)\rho-M(t)r\,,\label{eq:ric2}
\end{eqnarray}
\end{subequations}
with the parameters in it clarified in Figure \ref{fig:lambda}, was analytically solved for the case of weak depletion of the initial state, i.e., under the condition $|r|,\,|\rho|\ll 1$. Under this condition, we first dropped the term $M(t)\rho$ in Equation~(\ref{eq:ric1}), making this equation a quadratically nonlinear Riccati-type equation known from the two-level problem \cite{pluc, rost}. Next, the Riccati equation for $r$ was solved in the zero-order approximation by neglecting the term quadratic in $r$. A better solution to the Riccati equation for $r$, describing approximately the effect of nonlinearity in $r$, was then found in a perturbative way and this solution was substituted to Equation~(\ref{eq:ric2}). At a given $r$, Equation~(\ref{eq:ric2}) is a nonhomogeneous linear first-order differential equation with known formal solution \cite{parz}.

Now we extend our approach \cite{parz} to cover the case of significant depopulation of the initial state, particularly the spectacular case when the population is periodically exchanged between the two lower states lying in the bottoms of lambda, leaving the highest state practically unpopulated. Such a population exchange between states 1 and 3 was, in fact, obtained in the numerical calculations reported by both us (Figure~2(c) in \cite{parz} and the corresponding text) and Caldara and Fiordilino (Figure~2(b) in \cite{cal}). Specifically, an effective exchange between the mentioned states occurred when these states were separated by nearly the energy of two photons. This case of the two-photon separation of the lower states is the very one we focus our attention on throughout this paper. In terms of $r$ and $\rho$, the case to be considered means that $r$ is still small ($|r|\ll 1$) but $\rho$ can be large.

\section{The problem in terms of the Riccati equation}
\indent In Figure~\ref{fig:lambda}, we sketch the system under study. The initial state of the system has its eigenfrequency $\omega_1 = 0$. The parity $+(-)$ of the 1st state differs from that of the 2nd state and is the same as that of the 3rd state so the only non-zero dipole matrix elements in the system are $\mu_{12} = \langle 1|ez|2\rangle$ and $\mu_{23} = \langle 2|ez|3\rangle$. Instead of applying the variables $r$ and $\rho$ as in \cite{parz}, we now prefer to work with the equations for the population amplitudes $b_j$ \cite{parz}:
\begin{subequations}
\label{eq:set1} 
\begin{eqnarray}
i \dot{b}_1 &=& -\Omega(t)b_2\,,\label{eq:b1}\\
i \dot{b}_2 &=& \omega_{21}b_2 -\Omega(t)b_1-M(t)b_3\,,\label{eq:b2}\\
i \dot{b}_3 &=& \omega_{31}b_3 -M(t)b_2\,,\label{eq:b3}
\end{eqnarray}
\end{subequations}
where $\Omega(t)$ and $M(t)$ are the instantaneous Rabi frequencies for the electric-dipole $1\rightarrow 2$ and $2\rightarrow 3$ couplings, respectively, and $\omega_{21} = \omega_2$, $\omega_{31} = \omega_3$ are the eigenfrequencies of the other states. Precisely, $\Omega(t) = \Omega_R f(t) \cos(\omega_0 t)$ and $M(t) = M_R f(t) \cos(\omega_0 t)$, where $f(t)$ is a slowly varying pulse shape function and $\cos(\omega_0 t)$ refers to the laser-field oscillations as fast as the optical frequency $\omega_0\ll \omega_{21},\,\omega_{23}$. Moreover, $\Omega_R = \frac{\mu_{12}E_0}{\hbar}$ and $M_R = \frac{\mu_{23}E_0}{\hbar}$ are the familiar time-independent one-photon Rabi frequencies with $E_0$ being the peak electric-field strength.

Under the conditions $\omega_{21}\gg\omega_0$ and $|b_2|^2\ll 1$, we shall first reduce Equations~(\ref{eq:set1}) to an effective two-state set. Since the numerical calculations \cite{parz, cal} pointed to negligible population of level 2 when one meets nearly two-photon resonance between  states 1 and 3, we approximately eliminate $b_2$ from the set (Equations~(\ref{eq:set1})). First, we formally integrate Equation~(\ref{eq:b2}) in the range $\langle 0,t\rangle$. Since this equation alone is a nonhomogeneous linear in $b_2$ first-order differential equation, its formal solution is \cite{parz}
\begin{eqnarray}
b_2(t) = i e^{-i \omega_{21} t}\int_{0}^{t}(\Omega(t^{'})b_1(t^{'})+M(t^{'})b_3(t^{'}))e^{i\omega_{21}t^{'}}\ud t^{'}\,.\label{eq:solb2}
\end{eqnarray}
We know that $\Omega(0) = M(0) = 0$ and that (since $\omega_{21}\gg\omega_0$) these instantaneous Rabi frequencies oscillate slowly in time with respect to the function $e^{i\omega_{21}t}$. However, we a priori assume that the amplitudes $b_1$ and $b_3$ also change slowly in time as compared to $e^{i\omega_{21}t}$. Thus, the solutions for $b_1$ and $b_3$, if found under this adiabatic assumption, will have a limited range of applicability. This range will be defined later on. With this adiabatic assumption, the leading term in Equation~(\ref{eq:solb2}) is
\begin{eqnarray}
b_2(t) &\simeq& \frac{1}{\omega_{21}}\left(\Omega(t)b_1(t)+M(t)b_3(t)\right)\,.\label{eq:apb2}
\end{eqnarray}
This $b_2(t)$, when substituted to Equations~(\ref{eq:b1}) and (\ref{eq:b3}), gives a pair of equations for $b_1$ and $b_3$:
\begin{subequations}
\label{eq:set2}
\begin{eqnarray}
i\dot{b_1} &=& -\dot{\theta}b_1-\frac{\Omega(t)M(t)}{\omega_{21}}b_3\,,\\
i\dot{b_3} &=& \dot{\phi}b_3-\frac{\Omega(t)M(t)}{\omega_{21}}b_1\,,
\end{eqnarray}
\end{subequations}
where $\theta(t) = \int_0^t (\Omega^2(t^{'})/\omega_{21})\ud t^{'}$ and $\phi(t) = \int_0^t (\omega_{31}-(M^2(t^{'})/\omega_{21}))\ud t^{'}$. Above, $\dot{\theta}$ and $\dot{\phi}-\omega_{31}$ represent the instantaneous Stark shifts of states 1 and 3 caused by direct coupling of these states to state 2, while $\Omega(t)M(t)/\omega_{21}$ looks like an effective time-dependent coupling between states 1 and 3 through the intermediate state 2. Substituting $b_1=u\,\exp(i\theta(t))$ and $b_3 = v\,\exp(-i\phi(t))$, we transform set (\ref{eq:set2}) into
\begin{subequations}
\label{eq:set3}
\begin{eqnarray}
i\dot{u} &=& -S(t)v\,,\label{eq:u}\\
i\dot{v} &=& -S^{*}(t)u\,,\label{eq:v}
\end{eqnarray}
\end{subequations}
where
\begin{equation}
S(t) = \frac{\Omega(t)M(t)}{\omega_{21}}e^{-i\alpha(t)}
\label{eq:fullS}
\end{equation}
with the phase
\begin{eqnarray}
\alpha(t) &=& \theta(t)+\phi(t)\nonumber\\
&=& \omega_{31}t+\Delta_t+\Delta \int_0^t f^2(t^{'})\cos(2\omega_0 t^{'}) \ud t^{'}\nonumber\\
&\simeq & \omega_{31}t+\Delta_t+\frac{\Delta}{2\omega_0}f^2(t)\sin(2\omega_0 t)
\end{eqnarray}
depending on the optical field strength through the combined static Stark shift $\Delta = \frac{\Omega_R^2-M_R^2}{2\omega_{21}}$. Here, $\Delta_t = \Delta\int_0^t f^2(t^{'})\ud t^{'}$. When writing the final (approximate) form for the phase $\alpha(t)$ we have applied the condition that the pulse shape function $f(t)$ changes slowly with respect to $\cos(2\omega_0 t)$. Introducing a new variable $x = \frac{v}{u}$, set (\ref{eq:set3}) is then reduced to a single equation for this variable, namely the Riccati-type equation
\begin{eqnarray}
i\dot{x} &=&S(t)x^2-S^*(t)\,,\nonumber\\
         &=& (x^2-1)S(t)+2 i\,\mathrm{Im}(S(t))\,.\label{eq:riccati}
\end{eqnarray}
This equation has structurally the same form (but different $S(t)$) as the one met in the problem of a 2-level system \cite{pluc, rost}. The corresponding two-level system equation reads $i\,\dot{R} = S(t)R^2-S^{*}(t)$, where $R = (b_2/b_1)e^{i\omega_{21}t}$ and $S(t) = \Omega(t)e^{-i\omega_{21}t}$. In terms of the present $x$ and $S(t)$ one obtains from Equations~(\ref{eq:set3}) the required amplitudes:
\begin{subequations}
\label{eq:bric}
\begin{eqnarray}
b_1(t) &=& \exp\left[i\left(\int_0^t S(t^{'}) x(t^{'}) \ud t^{'}+\theta(t)\right)\right]\,,\\
b_3(t) &=& x\,\exp\left[i\left(\int_0^t S(t^{'}) x(t^{'}) \ud t^{'}-\phi(t)\right)\right]\,.
\end{eqnarray}
\end{subequations}
Finally, using the Fourier-Bessel expansion to the term $e^{i\alpha(t)}$, i.e., $\exp(i z \sin\phi) = \sum_{n = -\infty}^{\infty}J_n(z)\exp(in\phi)$ and also the relation $J_{n-1}(z)+J_{n+1}(z) = \frac{2 n}{z}J_n(z)$ satisfied by the Bessel functions $J_n(z)$ we write the present $S(t)$ defined by Equation~(\ref{eq:fullS}) as 
\begin{eqnarray}
S(t) &=& \frac{\Omega_R M_R}{2\omega_{21}}f^2(t)\sum_{n = -\infty}^{\infty}(-1)^n\label{eq:sapp1}\\ 
&\cdot &\left(1-n\frac{2\omega_0}{\Delta f^2(t)}\right)J_n\left(\frac{\Delta f^2(t)}{2\omega_0}\right)e^{-i[(\omega_{31}-2n\omega_0)t+\Delta_t]}\nonumber\\
&=& \sum_{n=-\infty}^{\infty}S_n(t)e^{-i[(\omega_{31}-2n\omega_0)t+\Delta_t]}\,.\nonumber
\end{eqnarray}
This $S(t)$ was derived taking fast oscillations of the electric field in the optical pulse in the form of $\cos(\omega_0 t)$. For the $\sin(\omega_0 t)$ oscillations one should replace $(-1)^n$ by $1$ in Equation~(\ref{eq:sapp1}). Equations~(\ref{eq:fullS}) and (\ref{eq:sapp1}) for $S(t)$ show a great role played by the term $\frac{\Delta}{2\omega_0}f^2(t)\sin(2\omega_0 t)$ in the phase $\alpha(t)$. First of all, it is a term allowing higher-order even-photon $1\rightarrow 3$ resonances in the system, i.e., higher than two-photon resonance. Moreover, this term introduces a complex field dependence to a fixed resonance because, due to this term, the field appears in the argument of the Bessel functions. In Section III, more details will be given on the role of this term in the phase $\alpha(t)$.

The key Riccati Equation~(\ref{eq:riccati}) is seen to play a crucial role in the above presentation. Obviously, it also results from Equations~(\ref{eq:ric}) for the variables $r$ and $\rho$. To show this, we first neglect the term quadratic in $r$ in Equation~(\ref{eq:ric1}) when considering the case of weak population of state 2. Then, the simplified Equation~(\ref{eq:ric1}) is solved in the adiabatic approximation with the result $r\simeq (\Omega(t)+M(t)\rho)/\omega_{21}$ analogous to Equation~(\ref{eq:apb2}). Substituting this $r$ to Equation~(\ref{eq:ric2}) and taking into account that $\rho = x e^{-i\alpha(t)}$ gives the Riccati Equation~(\ref{eq:riccati}).

\section{Approximate solution of the Riccati equation}
In \cite{parz, pluc, rost}, an analytic solution to the Riccati equation like Equation~(\ref{eq:riccati}) was given but for the case of weak depopulation of the initial state 1. That solution is not valid for the present case because now the initial-state population can completely be moved to the third state under the condition of two-photon $1\rightarrow 3$ resonance. The strong pumping of state 3 means intermediate and large values of $x^2$ and, thus, the dominant role of the first term on the right-hand side of Equation~(\ref{eq:riccati}). On the other hand, the second term on the right-hand side of Equation~(\ref{eq:riccati}) is expected to be non-negligible in the opposite limit of weak pumping of state 3, i.e., when $x^2$ is small. Mathematically, if
\begin{equation}
x_0(t) = i\tan\left(\int_0^t S(t^{'})\ud t^{'}\right)\,.
\end{equation}
is the solution of the reduced Equation~(\ref{eq:riccati}), $i\,\dot{x}_0 = (x_0^2-1)S(t)$, and
\begin{equation}
x_1(t) = 2\int_0^t\mathrm{Im}(S(t^{'}))\ud t^{'}
\end{equation}
solves the other reduced Equation~(\ref{eq:riccati}), $i\,\dot{x}_1 = 2 i\, \mathrm{Im}(S(t))$, then
\begin{eqnarray}
x(t) = x_0(t)+x_1(t)
\end{eqnarray}
is an approximate solution of the complete Riccati Equation~(\ref{eq:riccati}) under the condition $|(x_1^2+2x_0x_1)(S(\tau)/\omega_0)|\ll 1$, where $\tau = \omega_0 t$. Substituting the above $x(t)$ to Equations~(\ref{eq:bric}) results in
\begin{subequations}
\label{eq:bric2}
\begin{eqnarray}
b_1(t) &=& \cos\left(\int_0^t S(t^{'})\ud t^{'}\right) e^{i\theta(t)} \beta(t)\,,\\
b_3(t) &=& \left\{ i\sin\left(\int_0^t S(t^{'})\ud t^{'}\right)\right.\nonumber\\
    &+& \left.x_1\cos\left(\int_0^t S(t^{'})\ud t^{'}\right) \right\}e^{-i\phi(t)}\beta(t)\,,
\end{eqnarray}
\end{subequations}
where $\beta(t) = \exp\left[i\int_0^t  S(t^{'}) x_1(t^{'})\ud t^{'}\right]$.
In the following, we shall consider a special form of these amplitudes.

Let us assume that the pulse envelope function $f(t)$ rises from 0 to 1 in a short few-cycle time $t_{r}$ and then keeps constant value $f(t) = 1$ for $t>t_{r}$. Thus, for the rising time much shorter than the pulse length, we roughly approximate $\Delta_t \simeq \Delta t$. As a consequence, when the dynamic even-photon $1\rightarrow 3$ resonance takes place ($\omega_{31}+\Delta = P\omega_0$, where $P = 2n = 0,2,4,\dots$ if $\omega_{31}>0$; $P$ is negative if $\omega_{31}<0$), $S(t)$ of Equation~(\ref{eq:sapp1}) splits into its slowly and rapidly varying in time parts. The slow resonance part is that with $n = P/2$ only,

\begin{eqnarray}
S_{\frac{P}{2}}(t) &=& (-1)^{\frac{P}{2}}\frac{\Omega_R M_R}{2\omega_{21}}f^2(t)(1-\frac{P}{2}\frac{2\omega_0}{\Delta f^2(t)})J_{\frac{P}{2}}\left(\frac{\Delta f^2(t)}{2\omega_0}\right)\nonumber\\
&=&\Omega^{(P)}(t)\,, \label{eq:slow}
\end{eqnarray}
while the rest in $S(t)$ is the other part. This slow part will give dominant contribution to the integral $\int_0^t S(t^{'})\mathrm{d}t^{'}$ rooted as the argument in the trigonometric functions in Equations~(\ref{eq:bric2}). Thus, in analogy to the standard one-photon rotating-wave approximation (RWA) \cite{allen}, we now make generalized rotating-wave approximation (GRWA \cite{avet2002, gib2003,avet2008,parz2010}) in the case of a given even-photon $1\rightarrow 3$ resonance, i.e, neglect the rapidly oscillating terms in $S(t)$. In GRWA, $S(t) = \Omega^{(P)}(t)$ and is thus real entailing $x_1=0$, $\beta(t) = 1$ and $x =x_0 = i\tan\left(\int_0^t \Omega^{(P)}(t^{'})\ud t^{'}\right)$. The last $x$ is seen to be the exact solution of the Riccati Equation~(\ref{eq:riccati}) in its GRWA limit. The GRWA population amplitudes, obtained from Equations~(\ref{eq:bric2}), are thus
\begin{subequations}
\label{eq:solb}
\begin{eqnarray}
b_1(t) &=& \cos\left(\int_0^t \Omega^{(P)}(t^{'}) d t^{'}\right)e^{i\theta(t)}\,,\label{eq:solb1}\\
b_3(t) &=& i\sin\left(\int_0^t \Omega^{(P)}(t^{'}) d t^{'}\right)e^{-i\phi(t)}\,.\label{eq:solb3}
\end{eqnarray}
\end{subequations}
After the rising time $t_{r}$ has passed, the GRWA populations of levels 1 and 3 ($|b_j|^2$) are seen to oscillate at a constant $P$-photon Rabi frequency resulting from Equation~(\ref{eq:slow}) by putting $f(t) = 1$.

The applicability range of the above special solution (Equations~(\ref{eq:solb})) results from the requirement that $b_1$ and $b_3$ change slowly in time with respect to $\exp(i\omega_{21}t)$, where $\omega_{21}\gg\omega_0$. The amplitudes will change slowly if four inequalities are fulfilled: $|\omega_{31}/\omega_{21}|\ll 1$, $(\Omega_R/\omega_{21})^2\ll 1$, $(M_R/\omega_{21})^2\ll 1$ and $|\Omega^{(P)}/\omega_{21}|\ll 1$. The first inequality is realized in the lambda-configuration of the levels only (Figure~\ref{fig:lambda}), while all other inequalities impose the upper limit on the applied electric field. Since $\omega_{21}\gg\omega_0$, the strongest field is such that both $\Omega_R/\omega_0$ and $M_R/\omega_0$ are much smaller than $\omega_{21}/\omega_0$ allowing $\Omega_R/\omega_0$ and $M_R/\omega_0$ to be as high as of the order of 1. For this strongest field, one finds $|\Delta/2\omega_0|\ll 1$ justifying the approximation for the Bessel function $J_{P/2}\left(\frac{\Delta}{2\omega_0}f^2(t)\right)\simeq \frac{\left(\Delta f^2(t)/4\omega_0\right)^{P/2}}{(P/2)!}$. With this approximation, Equation~(\ref{eq:slow}) for $\Omega^{(P)}(t)$ reduces to:
\begin{subequations}
\label{eq:rabifreq}
\begin{eqnarray}
\Omega^{(0)}(t) &=& \frac{\Omega_R M_R}{2 \omega_{21}} f^2(t)\,,\label{eq:rabifreq1}\\
\Omega^{(P>0)}(t) &=& (-1)^q \Omega^{(0)}(t) \frac{\left(\frac{\Delta f^2(t)}{4\omega_0}\right)^q}{2(q!)}\,,\label{eq:rabifreq2}
\end{eqnarray}
\end{subequations}
where $q = \frac{P}{2}-1$.

Nearly the same solution as our Equations~(\ref{eq:solb}), but with $\Omega^{(P)}(t)$ given by Equations~(\ref{eq:rabifreq}) rather than general  Equation~(\ref{eq:slow}), can be obtained from the Floquet-like analysis. However, this analysis seems to be less comfortable and consumes much more time than our derivation. The basis for this analysis is the series expansion for the atomic amplitudes $b_1(t)$ and $b_3(t)$:
\begin{subequations}
\begin{eqnarray}
b_1(t) &=& \sum_{m=-\infty}^{\infty}\alpha_m(t)e^{-i 2 m \omega_0 t}\,,\\
b_3(t) &=& \sum_{m=-\infty}^{\infty}\beta_m(t)e^{-i 2 m \omega_0 t}\,.
\end{eqnarray}
\end{subequations}
When substituted to Equations~(\ref{eq:set2}), this expansion results in the evolutions for the Floquet-state amplitudes:
\begin{subequations}
\label{eq:floqueteq}
\begin{eqnarray}
i\,\dot{\alpha}_m &=& -(2 m \omega_0+\frac{\Omega_R^2}{2\omega_{21}}f^2(t))\alpha_m\nonumber\\
&-& \frac{\Omega_R^2}{4\omega_{21}}f^2(t)(\alpha_{m-1}+\alpha_{m+1})\\
&-& \frac{\Omega_R M_R}{4\omega_{21}}f^2(t)(2 \beta_m+\beta_{m-1}+\beta_{m+1})\nonumber\,,\\
i\,\dot{\beta}_m &=& (\omega_{31}-2 m \omega_0-\frac{M_R^2}{2\omega_{21}}f^2(t))\beta_m\nonumber\\
&-& \frac{M_R^2}{4\omega_{21}}f^2(t)(\beta_{m-1}+\beta_{m+1})\\
&-& \frac{\Omega_R M_R}{4\omega_{21}}f^2(t)(2 \alpha_m+\alpha_{m-1}+\alpha_{m+1})\nonumber\,.
\end{eqnarray}
\end{subequations}
Under the condition of $P = 2m$-photon resonance ($\omega_{31} - P\omega_0 = \delta_P\rightarrow 0$), Equations~(\ref{eq:floqueteq}) can be solved approximately in a truncated basis of states. The states of this basis are those forming the shortest coupling chain between the initial state of amplitude $\alpha_0$ and the final state of amplitude $\beta_{m = P/2}$. The number of intermediate states in this shortest chain depends on the multiphoton order $P$. For $P = 0$, i.e., the case of degeneracy ($\omega_{31} = \delta_P\rightarrow 0$) there is no intermediate state between the initial and final states characterized by $\alpha_0$ and $\beta_0$, respectively. There is also no intermediate state for $P=2$ but now the initial and final states are characterized by $\alpha_0$ and $\beta_1$, respectively. However, for $P = 2m\geq 4$ there are $P-2$ intermediate states characterized by $\alpha_1,\,\beta_1,\,\alpha_2,\,\beta_2,\dots,\,\alpha_{(P/2)-1},\,\beta_{(P/2)-1}$. By adiabatic elimination of all intermediate states one obtains a pair of equations for the amplitudes of the initial and final Floquet states coupled by $P$-photon resonance:
\begin{subequations}
\begin{eqnarray}
i\,\dot{\alpha}_0 &=& -\frac{\Omega_R^2}{2\omega_{21}}f^2(t) \alpha_0 - \Omega^{(P)}(t)\beta_{P/2}\,,\\
i\,\dot{\beta}_{P/2} &=& \left(\delta_P-\frac{M_R^2}{2\omega_{21}}f^2(t)\right) \beta_{P/2}\nonumber\\
&-& \Omega^{(P)}(t)\alpha_0\,,
\end{eqnarray}
\end{subequations}
where $\Omega^{(P)}(t)$ is given by Equations~(\ref{eq:rabifreq}). The last set looks structurally as Equations~(\ref{eq:set2}) and, under the conditions that led us to Equations~(\ref{eq:solb}), we have
\begin{subequations}
\label{eq:floquet}
\begin{eqnarray}
b_1(t) &\simeq & \alpha_0\nonumber\\
&=& \cos\left(\int_0^t \Omega^{(P)}(t^{'})\ud t^{'}\right)\nonumber\\
&\cdot & \exp\left[i\frac{\Omega_R^2}{2\omega_{21}}\int_0^t f^2(t^{'})\ud t^{'}\right]\,,\\
b_3(t) &\simeq & \beta_{P/2}e^{-i P \omega_0 t}\nonumber\\
&=& i\,\sin\left(\int_0^t \Omega^{(P)}(t^{'})\ud t^{'}\right)\nonumber\\
&\cdot & \exp\!\!\left[i\!\!\left(\!\!\Delta t - P\omega_0 t+\frac{M_R^2}{2\omega_{21}}\!\!\int_0^t\!\!\! f^2(t^{'})\ud t^{'}\!\!\right)\!\!\right].
\end{eqnarray}
\end{subequations}
The advantage of our Equations~(\ref{eq:solb}) is that they include the Rabi frequencies in their general form given by Equation~(\ref{eq:slow}), while the solution from the Floquet-like analysis (Equations~(\ref{eq:floquet})) includes the approximate Rabi frequencies given by Equations~(\ref{eq:rabifreq}).

\section{Results}

\begin{figure}[!ht]
\centering
\includegraphics[width=0.4\textwidth]{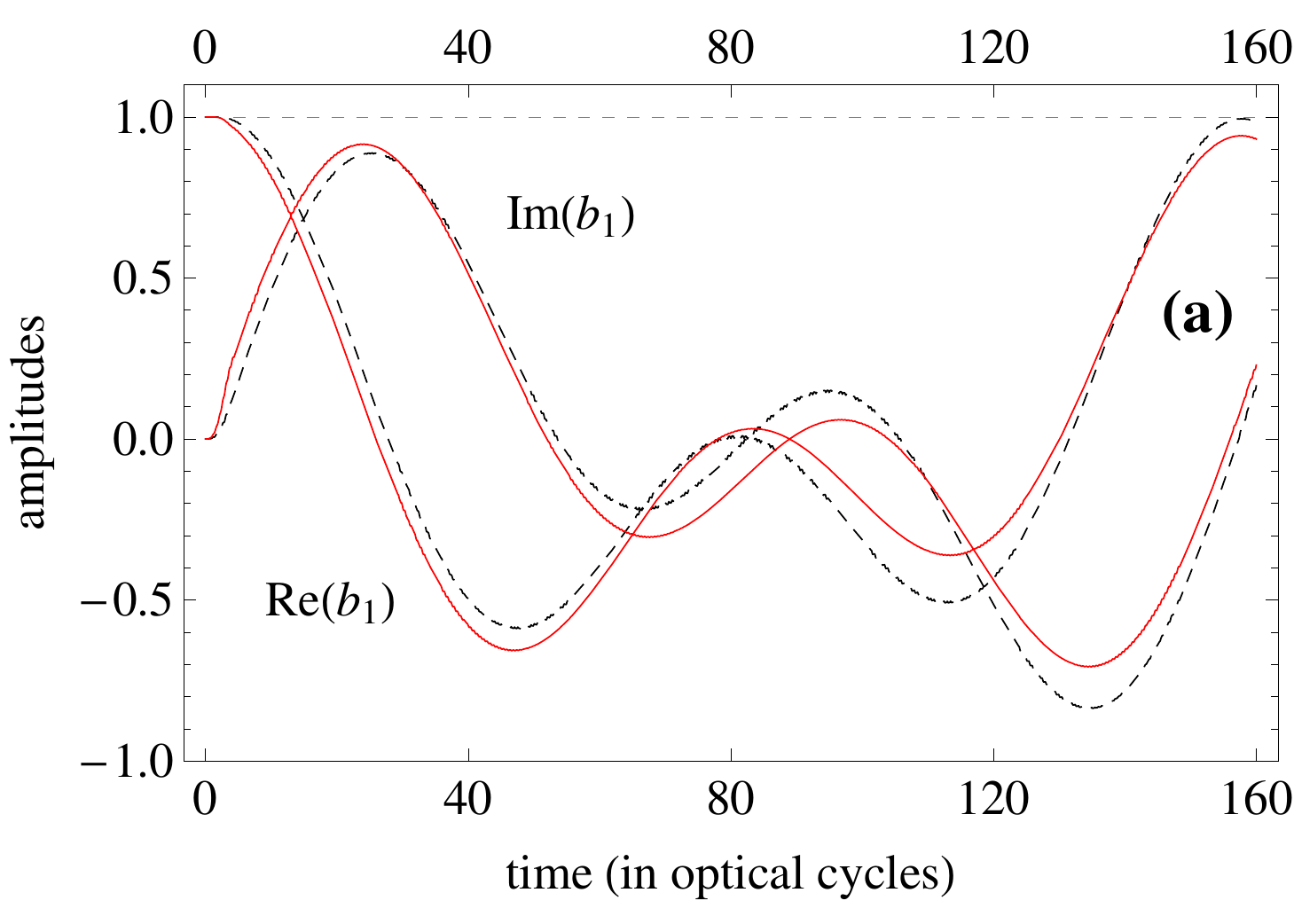}
\includegraphics[width=0.4\textwidth]{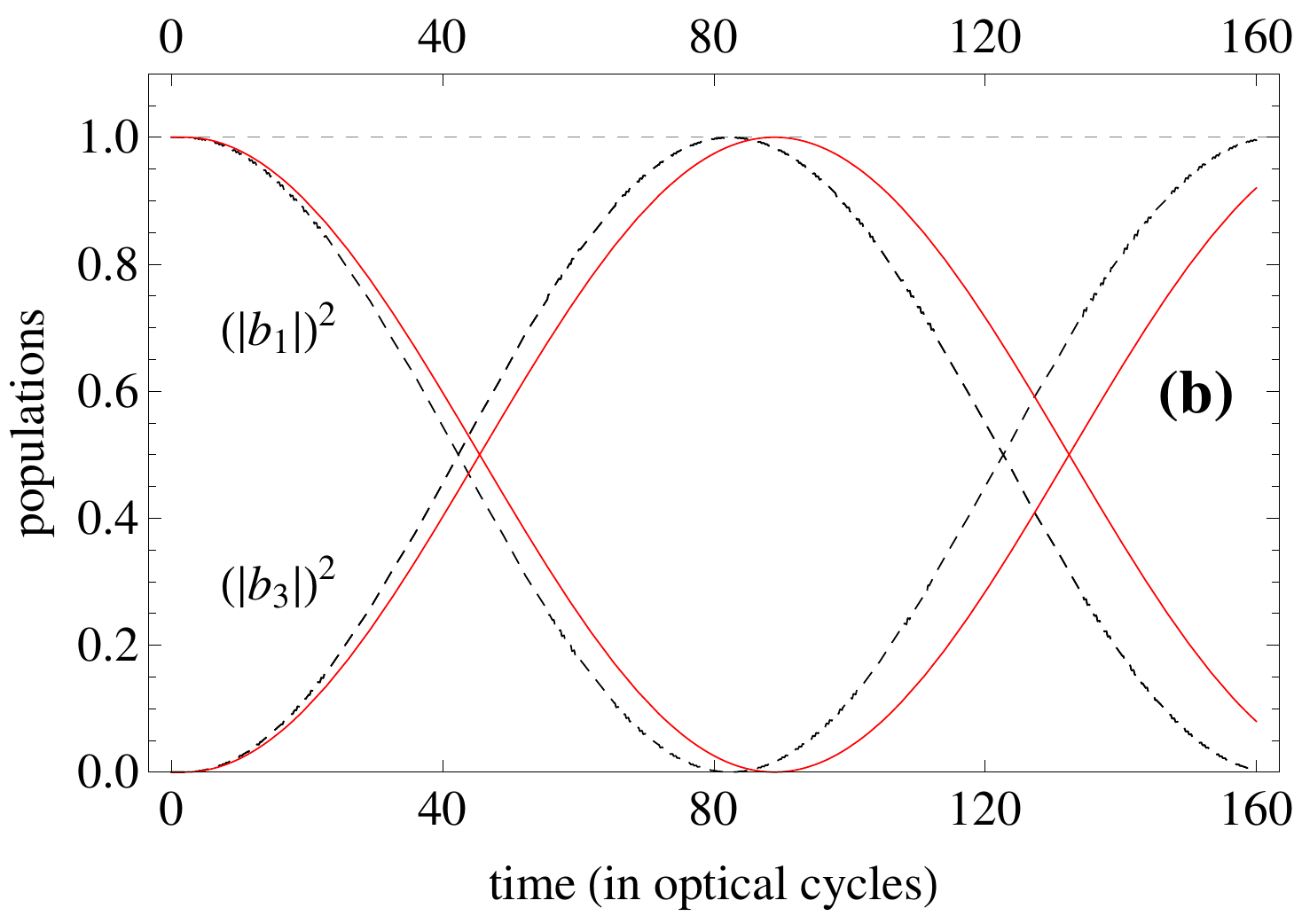}
\includegraphics[width=0.4\textwidth]{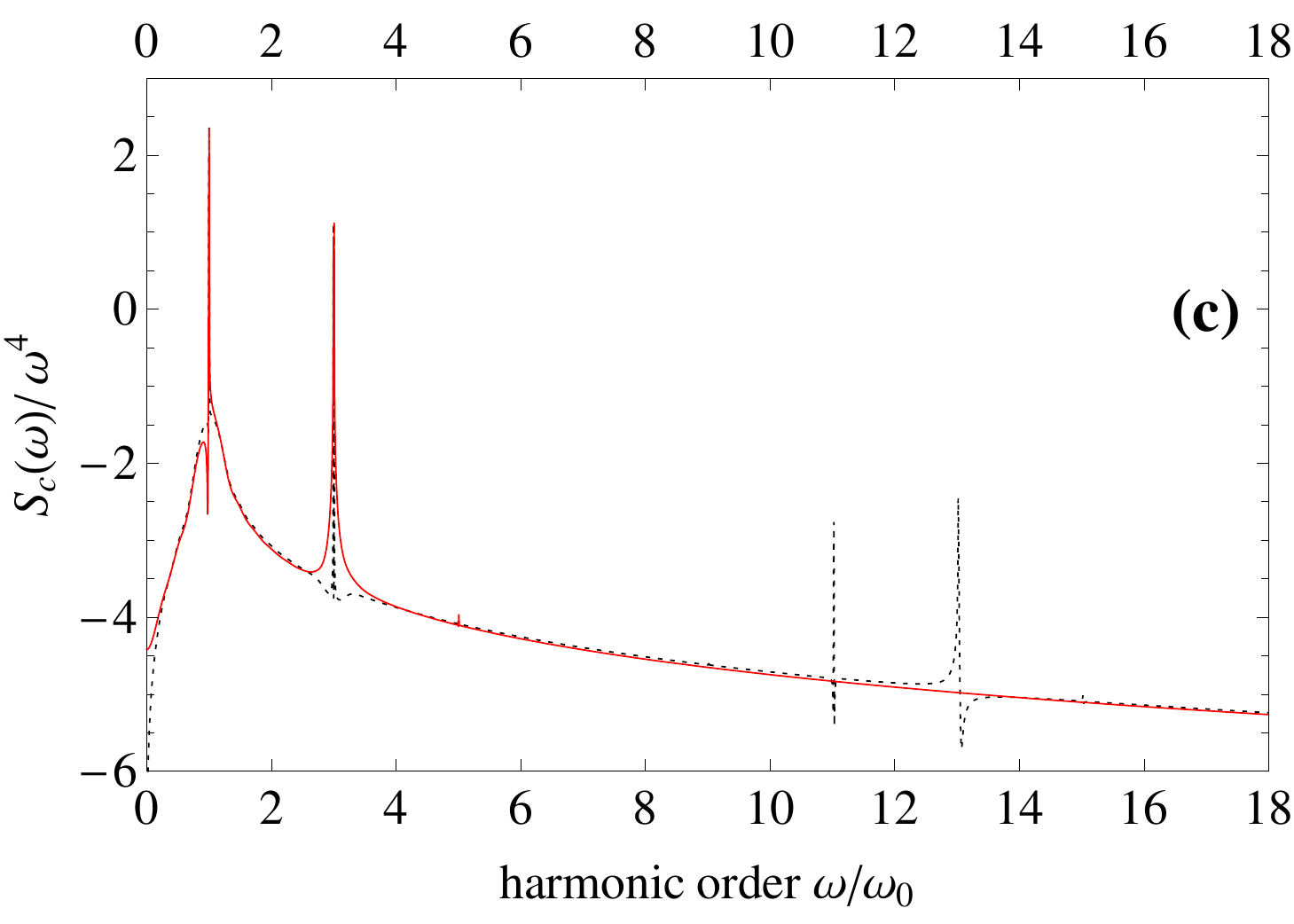}
\caption{(Color online) (a) Evolution of the real and imaginary parts of the initial-state population amplitude $b_1$ (we do not show the evolution of $b_3$ due to fast oscillations with frequency $2\omega_0$ under slowly changing envelope). (b) Two-photon Rabi oscillations of the populations in states 1 and 3. (c) Coherent spectrum of scattered light. The conditions: $\omega_{21}/\omega_0 = 13$, $\omega_{31}/\omega_0 = 1{.}99445$, $\Omega_{R}/\omega_0 = 0{.}5$, $M_R/\omega_0 = 0{.}3$, the pulse shape function $f(t)$ increases to its maximum value of 1 by four optical cycles and then keeps this value. Solid lines - the analytic results from Equations~(\ref{eq:solb}), dotted lines - the results of numerical integration of Equations~(\ref{eq:set1}).}\label{fig1}
\end{figure}
\begin{figure}[!ht]
\centering
\includegraphics[width=0.4\textwidth]{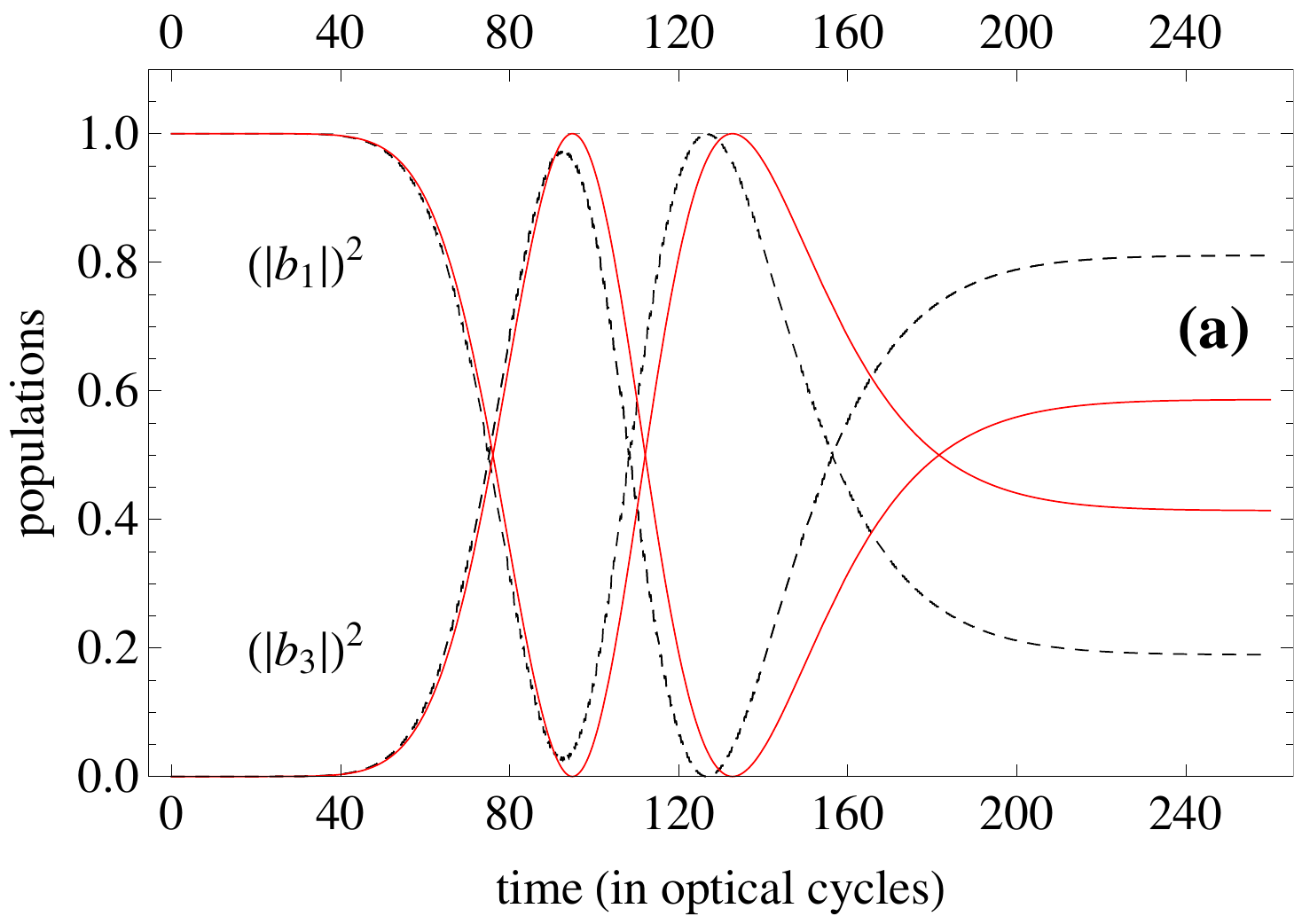}
\includegraphics[width=0.4\textwidth]{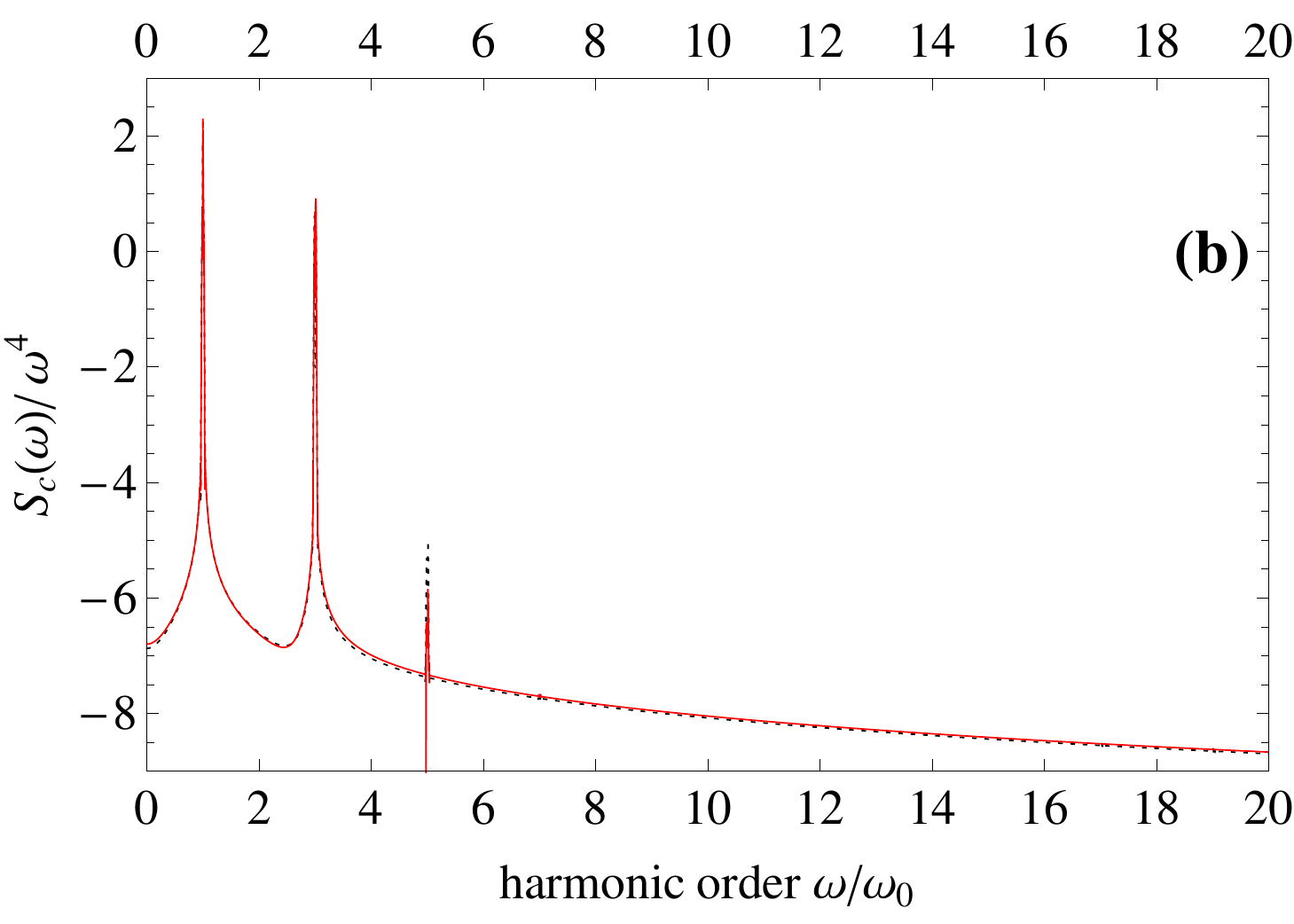}
\includegraphics[width=0.4\textwidth]{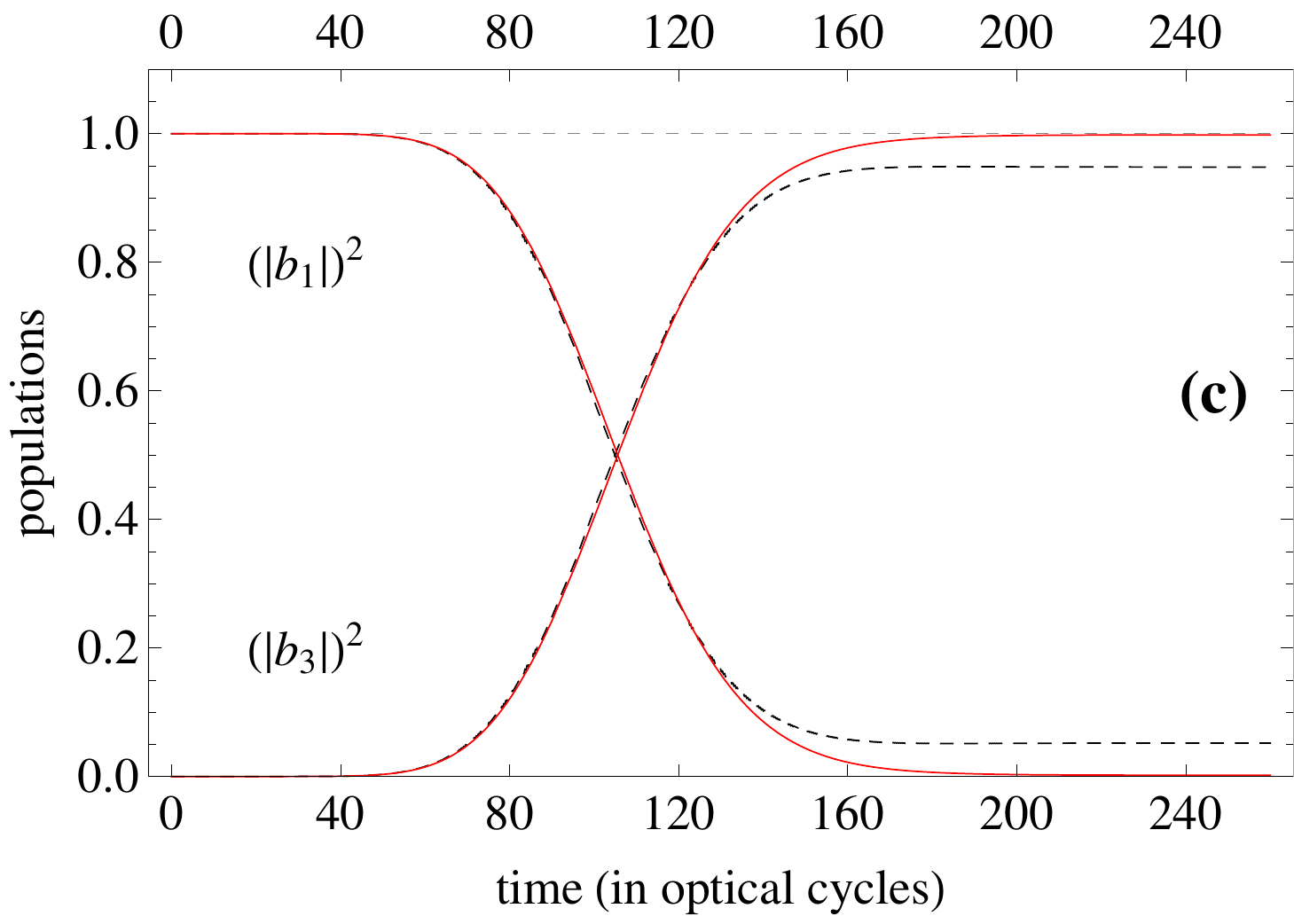}
\caption{(Color online) (a) Evolution of the populations in states 1 and 3. (b) Coherent spectrum of scattered light. The conditions as in \cite{cal}: $\omega_{21}/\omega_0 = 19$, $\omega_{23}/\omega_0 = 17$, $\omega_{31}/\omega_0 = 2$, $\Omega_{R}/\omega_0 = 0{.}8$, $M_R/\omega_0 = 0{.}7$, $f(t) = \left(\frac{t}{\tau}\right)^2\exp\left[1-\left(\frac{t}{\tau}\right)^2\right]$ with $t\geq 0$ and $\tau = 100\,T$. (c) As (a) but now for $\omega_{21}/\omega_0 = 50$, $\omega_{23}/\omega_0 = 48$ and the same all other parameters. Solid lines - the results after the adiabatic and generalized rotating-wave approximations, dotted lines - the result of numerical integration of Equations~(\ref{eq:set1}).}\label{fig:2}
\end{figure}
\indent In Figure~\ref{fig1}, we show the quality of the analytic GRWA population amplitudes (\ref{eq:solb}). We compare the results obtained from Equations~(\ref{eq:solb}) (solid lines) with those of the numerical integration (dotted lines) of the starting Equations~(\ref{eq:set1}). To ensure roughly the applicability of the analytic solution and the $1\rightarrow 3$ even-photon resonance we have taken the same parameters as in \cite{parz}, i.e., $\omega_{21}/\omega_0 = 13$,  $\omega_{31}/\omega_0 = 1.99445$,  $\Omega_R/\omega_0 = 0{.}5$, $M_R/\omega_0 = 0{.}3$ and, additionally, 4-cycle rise time of the pulse ($f(t) = \sin\left(\frac{\pi t}{2 t_{r}}\right)$ for $0\leq t\leq t_{r} = 4T = \frac{8\pi}{\omega_0}$ and $f(t) = 1$ for $t>t_{r}= 4T$). The assumed $\omega_{21}$, $\Omega_R$, and $M_R$ give $\Delta/\omega_0 = 0{.}00615$ so one meets the dynamic 2-photon $1\rightarrow 3$ resonance ($\omega_{31}+\Delta = 2{.}0006\omega_0$). For the above frequency-strength parameters the analytic amplitudes (\ref{eq:solb}) are reduced to the forms $b_1\simeq \cos\left(\Omega^{(2)}\int_0^t f^2(t^{'})\ud t^{'}\right)\cdot\exp\left(i\frac{\Omega_R^2}{2\omega_{21}}\int_0^t f^2(t^{'})\ud t^{'}\right)$ and  $b_3\simeq i \sin\left(\Omega^{(2)}\int_0^t f^2(t^{'})\ud t^{'}\right)\cdot\exp\left[i\left(-\omega_{31}t+\frac{M_R^2}{2\omega_{21}}\int_0^t f^2(t^{'})\ud t^{'}\right)\right]$, where $\Omega^{(2)} = \frac{\Omega_R M_R}{4\omega_{21}}$ results from the approximate Equation~(\ref{eq:rabifreq2}) for $\Omega^{(2)}(t)$. We have used these forms of $b_1$ and $b_3$ when drawing all analytic curves (solid lines) in Figure~\ref{fig1}. Figure~\ref{fig1}(a) shows the evolutions of the real and imaginary parts of $b_1$, Figure~\ref{fig1}(b) the 2-photon Rabi oscillations of populations ($|b_1|^2$,$|b_3|^2$) between levels 1 and 3 (the population of level 2 does not exceed $0{.}0015$ for the parameters chosen), and Figure~\ref{fig1}(c) the coherent part of the spectrum of light scattered by the system ($S_c(\omega)$ means squared modulus of the finite-time Fourier transform of the induced dipole in the system: $d_{induced} = 2\mu_{12}Re(b_1^* b_2)+2\mu_{23}Re(b_2^* b_3)$, where $b_2$ is to be calculated from Equation~(\ref{eq:apb2}) by applying the above $b_1$ and $b_3$). As seen, the analytic solution  explains well the trends observed in the results of numerical integration of Equations~(\ref{eq:set1}). Moreover, much better coincidence is now found between the analytic and numerical spectra in the low-frequency regime than in our previous paper (Figure 2(c) in \cite{parz}). This low-frequency part of spectrum includes three components. The strongest component simply corresponds to the elastic Rayleigh scattering. The two remaining components, $3\omega_0$ and $5\omega_0$, are the result of nonlinearity in the interaction between the system and light. As the field intensity increases (large $r,\,\Omega_R$ and $M_R$), the $5\omega_0$ component becomes more pronounced, as seen in Figure~\ref{fig:2}(b). One would expect the appearance of higher-order odd harmonics of $\omega_0$ in the spectrum as the intensity rises. However, we would like to mention that, in the high-frequency regime of the numerical spectrum in Figure~\ref{fig1}(c) ($\omega\simeq \omega_{21}$), one observes an additional low structure. The height of this structure is about five orders of magnitude lower than that of the leading peak in Figure~\ref{fig1}(c). Our analytic solution was not able to reproduce this additional low structure observed in the numerical spectrum. This is, probably, the cost we pay for the approximate procedures, particularly the procedure of adiabatic elimination of $b_2$ from Equations~(\ref{eq:set1}). The adiabatic elimination has turned out to be the main reason of quantitative differences between the solid and dotted curves in Figure~\ref{fig1}. It was found that this procedure works better for higher ratios $\omega_{21}/\omega_0$. By increasing this ratio well above the assumed value 13, the differences in the long-time population evolutions gradually decrease. Obviously, longer pulses are needed to observe the 2-photon Rabi oscillations when increasing the $\omega_{21}/\omega_0$ ratio.

In Figure~\ref{fig:2},  we present a different comparison for the conditions assumed by Caldara and Fiordilino \cite{cal} in their numerical integration of the Schr\"odinger equation. These conditions were: $\omega_{21}/\omega_0 = 19$, $\omega_{31}/\omega_0 = 2$, $\Omega_R/\omega_0 = 0{.}8$, $M_R/\omega_0 = 0{.}7$ and the  smooth pulse shape function $f(t) = \left(\frac{t}{\tau}\right)^2\exp\left[1-\left(\frac{t}{\tau}\right)^2\right]$, where $t\geq 0$ and $\tau = 100\,T$. For these conditions, Equations~(\ref{eq:slow}) and (\ref{eq:rabifreq}) for $\Omega^{(P)}(t)$ need to be multiplied by the factor $e^{-i\Delta_t}$. After including this factor we were not able to solve the Riccati Equation~(\ref{eq:riccati}) in an exact analytic way. So, we solved numerically the GRWA version of Equations~(\ref{eq:set3}) for $u$ and $v$ ($S(t)\rightarrow\Omega^{(P)}(t)$) and then obtained the solid-line curves shown in Figure~\ref{fig:2}. These curves, resulting from both the adiabatic and generalized rotating-wave approximations, are compared with the dot-line curves coming from the numerical integration of Equations~(\ref{eq:set1}). The comparison shows that, under the assumed conditions, the two approximations lead to results reflecting the main aspects of the exact (numerical) solution of Equations~(\ref{eq:set1}). Again, by increasing the ratio $\omega_{21}/\omega_0$ (from 19 to 50) much better agreement was found between the analytic and numerical curves (see Figure~\ref{fig:2}(c)). In the same time scale as in Figure~\ref{fig:2}(a), almost all the population is seen in Figure~\ref{fig:2}(c) to move from the initial state 1 to state 3 but without the return 2-photon transition from 3 to 1. The residual differences, still seen between the solid and dotted curves in Figure~\ref{fig:2}(c), may suggest that the approximate GRWA multiphoton Rabi frequencies given by Equations~(\ref{eq:rabifreq}) are not sufficiently accurate. However, taking $\Omega^{(P)}(t)$ in the form of Equation~(\ref{eq:slow}) has not removed the residual differences. As to the photon-emission spectrum shown in Figure~\ref{fig:2}(b), we do not observe now any low structure in the high-frequency regime ($\omega\simeq \omega_{21}$) of the numerical spectrum (dotted line). This makes a difference when compared to the numerical spectrum in Figure~\ref{fig1}(c).

\section{Summary}

With this paper we have extended the scope of validity of our recent analytic approach \cite{parz} to the problems of multiphoton population transfer in a three-level lambda-type system and scattering of low-frequency light by this system.The previous approach covered the case of weak depopulation of the initial state only. Now, the present version of the approach allows strong depopulation of the initial state as well. The amended approach was shown to explain the main numerical results, particularly the two-photon Rabi oscillations of population between the states forming the bottoms of $\Lambda$ and the dominant (if not all) peaks in the spectrum of scattered light. Though we considered the 2-photon resonance between the lower states in $\Lambda$, the approach holds for any even-photon resonance between the mentioned states as well as for the case of degeneration of these states.



\end{document}